\documentclass[conference]{IEEEtran}

\usepackage{cite}
\usepackage{amsmath,amssymb,amsfonts}
\usepackage{algorithmic}
\usepackage{graphicx}
\usepackage{textcomp}
\usepackage{xcolor}
\usepackage{epsfig}
\usepackage{float}
\usepackage{bm}
\usepackage{graphicx}
\usepackage{amsfonts,amssymb}
\usepackage{booktabs}
\usepackage{array}
\usepackage{tabularx}
\usepackage{multirow}
\usepackage{subfigure}
\usepackage{color}
\usepackage{epstopdf}
\usepackage{makecell}
\usepackage{url}
\usepackage{gensymb}
\usepackage{amsmath}
\usepackage{xcolor}
\usepackage[ruled]{algorithm2e}
\def\BibTeX{{\rm B\kern-.05em{\sc i\kern-.025em b}\kern-.08em
		T\kern-.1667em\lower.7ex\hbox{E}\kern-.125emX}}
\begin{document}
	
	\title{Cross-Block Difference Guided Fast CU Partition for VVC Intra Coding \vspace{-.5em}}

	\author{Hewei Liu$^{\ast}$, Shuyuan Zhu$^{\ast}$, Ruiqin Xiong$^{\dag}$, Guanghui Liu$^{\ast}$, and Bing Zeng$^{\ast}$\\
		
		{\begin{minipage}{\linewidth}\begin{center}
					\begin{tabular}{ccc}
						$^{\ast}$University of Electronic Science and Technology of China, Chengdu, China\\
						$^{\dag}$Peking University, Beijing, China\\
					\end{tabular}
		\end{center}\end{minipage}}
		\vspace{-1em}}
	
	
	\maketitle
%
%

\maketitle
%
%

	\begin{abstract}
		In this paper, we propose a new fast CU partition algorithm for VVC intra coding based on cross-block difference. This difference is measured by the gradient and the content of sub-blocks obtained from partition and is employed to guide the skipping of unnecessary horizontal and vertical partition modes. With this guidance, a fast determination of block partitions is accordingly achieved. Compared with VVC, our proposed method can save 41.64\% (on average) encoding time with only 0.97\% (on average) increase of BD-rate.
	\end{abstract}
	
	\begin{IEEEkeywords}
		VVC, intra coding, partition, mode, fast determination.
	\end{IEEEkeywords}
	
	\vspace{-0.5em}
	\section{Introduction}
	The versatile video coding (VVC)~\cite{1} adopts a number of advanced coding techniques. With these techniques, VVC achieves higher rate-distortion (R-D) performance than the high efficiency video coding (HEVC)~\cite{5}. However, the adoption of these techniques makes VVC suffer from higher complexity. Reducing the complexity of VVC is so necessary to make it more applicable.
	
	The QTMT structure used in the coding unit (CU) partition of VVC makes the block partition more flexible and content-adaptive, which offers considerable coding gain. Similar to the quadtree (QT) structure-based partition in HEVC, the QTMT-based one also determines the best CU partition by the R-D cost and the best mode is determined by the lowest R-D cost. Compared with the QT structure of HEVC, the QTMT structure introduces two new tree structures, i.e., the binary tree (BT) and the ternary tree (TT), which compose the multi-type tree (MTT) structure. Adopting MTT in VVC results in a rather complicated partition determination and dramatically increases the encoding time. Therefore, accelerating the determination of CU partition is one of the most effective solutions to reduce the complexity for VVC.
	
	Over the past few years, both the learning-based and  non-learning-based methods have been proposed to implement the fast CU partition for VVC intra coding. In the learning-based methods, the shape-adaptive convolutional neural network (CNN) was employed to rapidly process various CU partition shapes for VVC~\cite{6}. A low-complexity coding tree unit (CTU) structure decision algorithm was proposed in~\cite{7}, where the CU partition was modeled as a multi-classification problem and both the texture information and context information were used as the classification features. The support vector machine was also introduced in the fast prediction of the CU splitting direction~\cite{8}. Meanwhile, the Bayesian rule-based classifier was employed to achieve the fast CU partition in~\cite{9} which used the partition types and intra prediction modes of children CUs as the classification features.
	Although the learning-based solution is popular in the design of the fast CU partition methods, these proposed methods either suffer from the heavy training workload or rely on the accuracy of classifier, which limits their applications. Compared with the learning-based methods, the non-learning-based ones are normally implemented with lower workload based on the texture information of video frame. For instance, the block variance and gradient were utilized to accelerate the rectangular partition~\cite{10}.
	
	In this work, we propose a fast CU partition method for VVC intra coding based on the cross-block difference which is measured by the gradient and the content of sub-blocks obtained from partition. With our method, the unnecessary horizontal and vertical partitions of both the BT and TT structures can be effectively skipped without the R-D cost-based determination.
	
	\section{Relationship between block partition and cross-block difference} \label{Sec.relationship}
	
	\subsection{Background}
	The QTMT structure is introduced in VVC to achieve a fine block partition. There are six splitting modes in the QTMT-based partition. The first four modes are binary tree horizontal splitting (BHS), binary tree vertical splitting (BVS), ternary tree horizontal splitting (THS) and ternary tree vertical splitting (TVS). Besides these modes, there are two additional modes, i.e., quadtree splitting (QTS) and no splitting (NOS).

	%
	
	
	The previous works~\cite{9,10} demonstrate that the block partition is related to the texture characteristic of block. The texture information can be employed to guide the partition targeting a low complexity. More specifically, the smooth and homogeneous areas should be split into blocks with larger sizes, whereas the areas with complicated textures are usually split into blocks with smaller sizes.
	
	In this work, we aim to achieve a fast mode determination by skipping the horizontal and vertical partitions based on the difference between sub-blocks obtained from a split block. With four horizontal and vertical partition modes in VVC, i.e., BHS, BVS, THS and TVS, we can separate a block into two or three sub-blocks which have different shapes as shown in Fig.\ref{sp}. To compare the difference between sub-blocks, we define the cross-block difference as
	
	\vspace{-.5em}
	\begin{equation} \label{Eq.1}
	M(p_1,p_2) = \max (p_1,p_2)/\min (p_1,p_2),
	\end{equation}
	
	\noindent
	where $p_1$ and $p_2$ are measurement parameters for two compared sub-blocks. Note that $M(p_1,p_2)$ approximates $1$ if $p_1$ and $p_2$ are close enough, while $M(p_1,p_2)$ will be large if $p_1$ and $p_2$ are quite different. In our work, we utilize $M(p_1,p_2)$ to determine which partition modes should be skipped based on the characteristic of sub-blocks.
	
	\subsection{Gradient difference between sub-blocks}
	We firstly study the relationship between the partition strategy and the sub-block gradient. Given a sub-block $\mathbf{A}$, we can obtain its four directional gradients, i.e., the horizontal gradient $G^{(h)}$, vertical gradient $G^{(v)}$, diagonal gradient ${G^{(d)}}$ and anti-diagonal gradient ${G^{(a)}}$, by performing the Sobel operator \cite{11} on it as
	
	\vspace{-1em}
	\begin{eqnarray}\label{Eq.0}
	\setlength{\arraycolsep}{6pt}
	G^{(dir)}=
	{S^{(dir)}}
	\circledast\mathbf{A}, dir = h,v,d,a.
	\end{eqnarray}
	
	\noindent
	where $\circledast$ denotes the convolution operation and $S$ is the corresponding Sobel operator to obtain a directional gradient.
	
	\begin{figure} [t]
		\begin{minipage}[a]{0.48\linewidth}
			\centering
			\centerline{\epsfig{figure=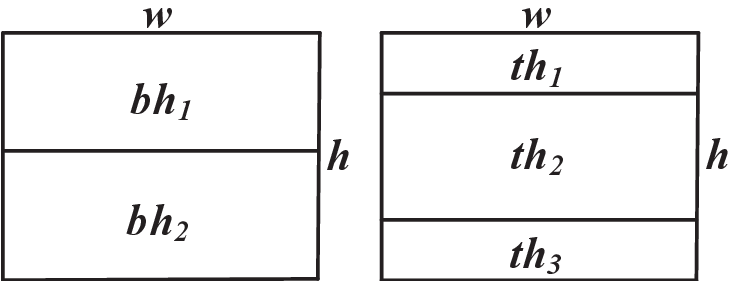,height=1.8cm,width=3.8cm}}
			\centerline{\small(a)}\medskip
		\end{minipage}
		\hfill
		\begin{minipage}[a]{0.48\linewidth}
			\centering
			\centerline{\epsfig{figure=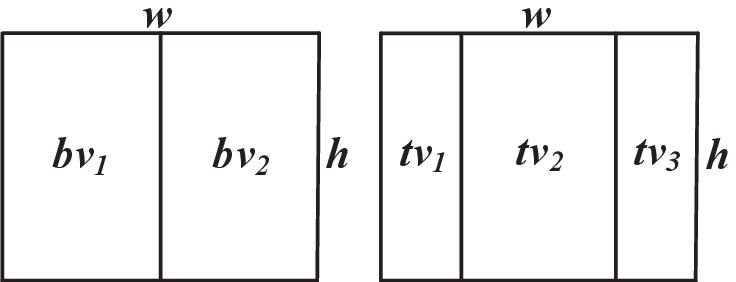,height=1.8cm,width=3.8cm}}
			\centerline{\small(b)}\medskip
		\end{minipage}
		\small{\caption{The sub-blocks obtained by separating a block with binary tree and ternary tree. (a) Horizontal sub-blocks. (b) Vertical sub-blocks.}\label{sp}}
		\vspace{-1em}
	\end{figure}

	
	We study the relationship between the partition strategy and the sub-block gradient based on some coded frames. Specifically, we select four video sequences from class D, including \emph{BasketballPass}, \emph{BlowingBubbles}, \emph{BQSquare} and \emph{RaceHorses}, and encode 50 frames of each sequence with the all-intra (AI) configuration. Four coding scenarios are adopted in our experiment, where the quantization parameters (QPs) are specified as $\{22, 27, 32, 37\}$, respectively. For all the scenarios and all the encoded frames, we compare the difference of sub-blocks based on their directional gradients. For each sub-block, the average gradient along a specific direction over all the pixels is utilized as its directional gradient. According to Fig. \ref{sp}, we use the notations listed in Table \ref{notations} to denote four directional gradients for a sub-block which obtained by the horizontal separation (HS) or vertical separation (VS) of a block with binary tree or ternary tree.
	
	To obtain the optimal partition mode for a block, all the candidate modes will be attempted and the one with the lowest R-D cost will be selected as the optimal mode. Besides the optimal partition, we can also separate each block into two or three sub-blocks along the horizontal or vertical directions with binary tree or ternary tree, which is like the BHS, BVS, THS and TVS-based partition. For each separation, we calculate the directional gradient differences between the sub-blocks and present the corresponding results in Table \ref{HGR}-\ref{135GR}. According to the results presented in Table \ref{HGR}-\ref{135GR}, it is found that splitting a block along a specific direction is highly related to the gradient differences of the resulting sub-blocks. For example, the sub-blocks obtained by the BHS-based partition always have bigger $M(G_{{bh}_1}^{(h)},G_{{bh}_2}^{(h)})$, $M(G_{{bh}_1}^{(v)},G_{{bh}_2}^{(v)})$, $M(G_{{bh}_1}^{(d)},G_{{bh}_2}^{(d)})$ and $M(G_{{bh}_1}^{(a)},G_{{bh}_2}^{(a)})$ than the ones obtained by the partition of other modes. These results indicate that we may determine an effective partition for a block based on the gradient differences of the sub-blocks.
	
	\begin{table}[t]
		\renewcommand\arraystretch{1.2}
		\centering
		\fontsize{7}{7}\selectfont
		\caption{Notations of gradients for horizonal and vertical separations}\label{notations}
		\vspace{-.5em}
		\begin{tabular}
			{c|c|c|c|c}
			\Xhline{1.2pt}
			\hline \hline
			&Horizontal&Vertical&Diagonal&Anti-diagonal\\ \hline
			BT-based HS &$G_{{bh}_i}^{(h)}$&$G_{{bh}_i}^{(v)}$&$G_{{bh}_i}^{(d)}$&$G_{{bh}_i}^{(a)}$\\ \hline
			TT-based HS&$G_{{th}_i}^{(h)}$&$G_{{th}_i}^{(v)}$&$G_{{th}_i}^{(d)}$&$G_{{th}_i}^{(a)}$\\ \hline
			BT-based VS&$G_{{bv}_i}^{(h)}$&$G_{{bv}_i}^{(v)}$&$G_{{bv}_i}^{(d)}$&$G_{{bv}_i}^{(a)}$\\ \hline
			TT-based VS&$G_{{tv}_i}^{(h)}$&$G_{{tv}_i}^{(v)}$&$G_{{tv}_i}^{(d)}$&$G_{{tv}_i}^{(a)}$\\ \hline\hline
			\Xhline{1.2pt}
		\end{tabular}
		\vspace{-1em}
	\end{table}
	
	\subsection{Content difference between sub-blocks}
	Compared with the BT-based splitting, fewer TT-based partition modes will be attempted in practical VVC coding\cite{12}. Therefore, we intend to develop a more accurate partition for the BT-based mode by with the guidance of both the gradient difference and the content difference between sub-blocks. We evaluate content difference between a $w_1 \times h_1$ sub-block $b_1$ and a $w_2 \times h_2$ sub-block $b_2$ as
	
	\vspace{-1em}
	\begin{eqnarray} \label{Eq.4}
	\begin{aligned}
	D({b_1},{b_2})= \frac{1}{w_1 \times h_1}& {\sum\limits_{x = 0}^{w_1-1}} {\sum\limits_{y = 0}^{h_1-1}} \Big({b_1(x,y)}-\\
	&\big(\frac{1}{{w_2 \times h_2}}{\sum\limits_{x = 0}^{w_2-1}} {\sum\limits_{y = 0}^{h_2-1} {{b_2(x,y)}}}\big) \Big)^2.
	\end{aligned}
	\end{eqnarray}
	
	According to Eq. (\ref{Eq.4}), when the contents of $b_1$ and $b_2$ are similar, the average pixel-levels of them will also be similar and their difference will be small. It indicates that the employed splitting mode generates sub-blocks with similar contents. As a result, this mode should be skipped.
	
	\section{Our Proposed Method}
	
	In this section, we propose a fast CU partition scheme for VVC intra coding by skipping the unnecessary horizontal and vertical partitions based on the cross-block difference.
	
	\subsection{Fast determination for the binary tree-based partition}
	Our previous results in Section \ref{Sec.relationship} reveals that the cross-block difference, especially the gradient difference, is highly related to the partition mode. Inspired by this fact, we define the gradient-based difference flags, i.e., ${d}^{(g)}_{bh}$ and ${d}^{(g)}_{bv}$, and the content-based difference flags, i.e., ${d}^{(c)}_{bh}$ and ${d}^{(c)}_{bv}$, to determine the mode skipping. Specifically, ${d}^{(g)}_{bh}$ and ${d}^{(g)}_{bv}$ are defined as
	
	\vspace{-.5em}
	\begin{eqnarray} \label{Eq.2}
	{d}^{(g)}_{bh}=
	\left\{
	\begin{array}{rcl}
	\begin{aligned}
	&0, if {\left(\begin{subarray}\\
		{M(G_{bh_1}^{(h)},G_{bh_2}^{(h)}) < T_1}\\
		\&\&{M(G_{bh_1}^{(v)},G_{bh_2}^{(v)})< T_1}\\
		\&\&{M(G_{bh_1}^{(d)}, G_{bh_2}^{(d)}) < T_1}\\
		\&\&{M(G_{bh_1}^{(a)},G_{bh_2}^{(a)}) < T_1}
		\end{subarray}\right)}\\
	&1, otherwise\\
	\end{aligned}
	\end{array} \right.
	\end{eqnarray}

	\noindent
	and
	
	\vspace{-1em}
	\begin{eqnarray} \label{Eq.3}
	{d}^{(g)}_{bv}=
	\left\{
	\begin{array}{rcl}
	\begin{aligned}
	&0, if {\left(\begin{subarray}\\
		{M(G_{bv_1}^{(h)},G_{bv_2}^{(h)}) < T_1}\\
		\&\&{M(G_{bv_1}^{(v)},G_{bv_2}^{(v)})< T_1}\\
		\&\&{M(G_{bv_1}^{(d)}, G_{bv_2}^{(d)}) < T_1}\\
		\&\&{M(G_{bv_1}^{(a)},G_{bv_2}^{(a)}) < T_1}
		\end{subarray}\right)}\\
	&1, otherwise\\
	\end{aligned}
	\end{array}, \right.
	\end{eqnarray}
	
	\noindent
	where $T_1$ is the pre-defined threshold. Note that ${d}^{(g)}_{bh}=0$ or ${d}^{(g)}_{bv}=0$ indicates that the resulting sub-blocks split by using the BHS mode or the BVS mode will generate two similar sub-blocks. Therefore, the corresponding partition mode should be skipped. When ${d}^{(g)}_{bh}=1$ or ${d}^{(g)}_{bv}=1$, we will further exam the content difference to refine the determination. 
	
	%
	
	To use content difference guide the partition, we define the content-based difference flags based on Eqs. (\ref{Eq.1}) and (\ref{Eq.4}) as
	
	\vspace{-.5em}
	\begin{eqnarray} \label{Eq.5}
	{d}^{(c)}_{bh}=
	\left\{
	\begin{array}{rcl}
	\begin{aligned}
	&1, if {\left(\begin{subarray}\\
		\frac{M(D(bv_1,bv_2),D(bv_2,bv_1))}{M(D(bh_1,bh_2),D(bh_2,bh_1))}< 1\\
		\& \& M(D(bh_1,bh_2),D(bh_2,bh_1)) > T_2
		\end{subarray}\right)}\\
	&0, otherwise\\
	\end{aligned}
	\end{array} \right.
	\end{eqnarray}
	
	\noindent
	and
	
	\vspace{-1em}
	\begin{eqnarray} \label{Eq.6}
	{d}^{(c)}_{bv}=
	\left\{
	\begin{array}{rcl}
	\begin{aligned}
	&1, if {\left(\begin{subarray}\\
		\frac{M(D(bh_1,bh_2),D(bh_2,bh_1))}{M(D(bv_1,bv_2),D(bv_2,bv_1))}< 1\\
		\& \& M(D(bv_1,bv_2),D(bv_2,bv_1)) > T_2
		\end{subarray}\right)}\\
	&0, otherwise\\
	\end{aligned}
	\end{array}, \right.
	\end{eqnarray}
	
	\noindent
	where $T_2$ is the pre-defined threshold. According to Eqs. (\ref{Eq.5}) and (\ref{Eq.6}), if $bh_1$ and $bh_2$ are different enough, i.e., ${d}^{(c)}_{bh}=1$, the block prefers to be split horizontally and the vertical splitting is accordingly skipped. Similarly, if $bv_1$ and $bv_2$ are different enough, i.e., ${d}^{(c)}_{bv}=1$, splitting  block along the vertical direction should be more effective and the horizontal separation should be skipped.
	
	We summary our binary tree-based partition algorithm in Algorithm \ref{Alg.1}, where ${skip}_{bh}=1$ and ${skip}_{bv}=1$ indicate skipping the BHS mode and BVS mode, respectively.
	
	\begin{table}[t]
		\vspace{-.5em}
		\renewcommand\arraystretch{1.2}
		\centering
		\fontsize{7}{7}\selectfont
		\caption{Gradient differences of two sub-blocks obtained by horizontal separation}\label{HGR}
		\vspace{-.5em}
		\begin{tabular}{c|p{15pt}p{15pt}p{15pt}p{15pt}p{15pt}p{15pt}}
			\Xhline{1.2pt}
			\hline \hline
			& BHS & THS& BVS & TVS & QTS & NOS\\\hline
			$M(G_{{bh}_1}^{(h)},G_{{bh}_2}^{(h)})$ &\textbf{2.51}&2.40&1.75&1.60&1.82&2.02\\ \hline
			$M(G_{{bh}_1}^{(v)},G_{{bh}_2}^{(v)})$ &{\textbf{2.81}}&2.46&2.00&1.82&1.89&2.19\\ \hline
			$M(G_{{bh}_1}^{(d)},G_{{bh}_2}^{(d)})$ &{\textbf{2.52}}&2.30&1.76&1.63&1.75&2.05\\ \hline
			$M(G_{{bh}_1}^{(a)},G_{{bh}_2}^{(a)})$ &{\textbf{2.56}}&2.38&1.78&1.65&1.77&2.06\\ \hline\hline
			
			\Xhline{1.2pt}
		\end{tabular}
		\vspace{-.5em}
	\end{table}
	
	\begin{table}[t]
		\renewcommand\arraystretch{1.2}
		\centering
		\fontsize{7}{7}\selectfont
		\caption{Gradient differences of two sub-blocks obtained by vertical separation} \label{45GR}
		\vspace{-.5em}
		\begin{tabular}{c|p{15pt}p{15pt}p{15pt}p{15pt}p{15pt}p{15pt}}
			\Xhline{1.2pt}
			\hline \hline
			& BHS & THS & BVS  & TVS & QTS & NOS\\ \hline
			$M(G_{{bv}_1}^{(h)},G_{{bv}_2}^{(h)})$&2.18&2.03&{\textbf{3.20}}&2.90&2.50&2.29\\ \hline
			$M(G_{{bv}_1}^{(v)},G_{{bv}_2}^{(v)})$ &1.72&1.59&{\textbf{2.40}}&2.22&1.83&1.93\\ \hline	
			$M(G_{{bv}_1}^{(d)},G_{{bv}_2}^{(d)})$ &1.83&1.69&{\textbf{2.65}}&2.40&1.99&2.04\\ \hline
			$M(G_{{bv}_1}^{(a)},G_{{bv}_2}^{(a)})$ &1.82&1.69&{\textbf{2.64}}&2.41&2.01&2.03\\ \hline\hline
			
			\Xhline{1.2pt}
		\end{tabular}
		\vspace{-1em}
	\end{table}
	
	\begin{table}[t]
		\vspace{-.5em}
		\renewcommand\arraystretch{1.2}
		\centering
		\fontsize{7}{7}\selectfont
		\caption{Gradient differences of three sub-blocks obtained by horizontal separation} \label{VGR}
		\vspace{-.5em}
		\begin{tabular}{c|p{15pt}p{15pt}p{15pt}p{15pt}p{15pt}p{15pt}}
			\Xhline{1.2pt}
			\hline \hline
			& BHS & THS & BVS  & TVS & QTS & NOS\\ \hline
			$M(G_{{th}_1}^{(h)},G_{{th}_2}^{(h)})$&2.27&{\textbf{2.49}}&1.72&1.60&1.72&1.74 \\ \hline
			$M(G_{{th}_1}^{(h)},G_{{th}_3}^{(h)})$&3.77&{\textbf{3.88}}&2.43&2.16&2.51&2.39\\ \hline
			$M(G_{{th}_2}^{(h)},G_{{th}_3}^{(h)})$&2.21&{\textbf{2.39}}&1.67&1.59&1.69&1.77\\ \hline
			$M(G_{{th}_1}^{(v)},G_{{th}_2}^{(v)})$ &2.57&{\textbf{2.69}}&1.98&1.85&1.92&1.94\\ \hline
			$M(G_{{th}_1}^{(v)},G_{{th}_3}^{(v)})$ &3.70&{\textbf{3.99}}&2.63&2.38&2.58&2.53\\ \hline
			$M(G_{{th}_2}^{(v)},G_{{th}_3}^{(v)})$ &2.54&{\textbf{2.70}}&1.97&1.84&1.95&1.95\\ \hline
			$M(G_{{th}_1}^{(d)},G_{{th}_2}^{(d)})$ &2.31&{\textbf{2.44}}&1.74&1.61&1.71&1.80\\ \hline
			$M(G_{{th}_1}^{(d)},G_{{th}_3}^{(d)})$ &3.52&{\textbf{3.59}}&2.32&2.10&2.36&2.40\\ \hline
			$M(G_{{th}_2}^{(d)},G_{{th}_3}^{(d)})$ &2.30&{\textbf{2.35}}&1.68&1.60&1.72&1.79\\ \hline
			$M(G_{{th}_1}^{(a)},G_{{th}_2}^{(a)})$ &2.35&{\textbf{2.48}}&1.76&1.62&1.73&1.79\\ \hline
			$M(G_{{th}_1}^{(a)},G_{{th}_3}^{(a)})$ &3.60&{\textbf{3.79}}&2.38&2.16&2.41&2.41\\ \hline
			$M(G_{{th}_2}^{(a)},G_{{th}_3}^{(a)})$ &2.31&{\textbf{2.48}}&1.72&1.64&1.76&1.83\\ \hline\hline
			
			\Xhline{1.2pt}
		\end{tabular}
	\end{table}
	
	\begin{table}[t]
		\vspace{-1em}
		\renewcommand\arraystretch{1.2}
		\centering
		\fontsize{7}{7}\selectfont
		\caption{Gradient differences of three sub-blocks obtained by vertical separation} \label{135GR}
		\vspace{-.5em}
		\begin{tabular}{c|p{15pt}p{15pt}p{15pt}p{15pt}p{15pt}p{15pt}}
			\Xhline{1.2pt}
			\hline \hline
			& BHS & THS & BVS  & TVS & QTS & NOS\\ \hline
			$M(G_{{tv}_1}^{(h)},G_{{tv}_2}^{(h)})$&2.19&2.11&2.95&{\textbf{3.23}}&2.38&2.06\\ \hline
			$M(G_{{tv}_1}^{(h)},G_{{tv}_3}^{(h)})$&3.07&2.98&4.55&{\textbf{4.94}}&3.78&2.84\\ \hline
			$M(G_{{tv}_2}^{(h)},G_{{tv}_3}^{(h)})$&2.07&2.04&2.81&{\textbf{2.95}}&2.46&1.96\\ \hline
			$M(G_{{tv}_1}^{(v)},G_{{tv}_2}^{(v)})$ &1.67&1.56&2.07&{\textbf{2.31}}&1.72&1.76\\ \hline
			$M(G_{{tv}_1}^{(v)},G_{{tv}_3}^{(v)})$ &2.29&2.12&3.35&{\textbf{3.48}}&2.48&2.39\\ \hline
			$M(G_{{tv}_2}^{(v)},G_{{tv}_3}^{(v)})$ &1.63&1.57&2.09&{\textbf{2.25}}&1.73&1.71\\ \hline
			$M(G_{{tv}_1}^{(d)},G_{{tv}_2}^{(d)})$ &1.78&1.71&2.34&{\textbf{2.60}}&1.91&1.85\\ \hline
			$M(G_{{tv}_1}^{(d)},G_{{tv}_3}^{(d)})$ &2.39&2.27&3.56&{\textbf{3.73}}&2.74&2.49\\ \hline
			$M(G_{{tv}_2}^{(d)},G_{{tv}_3}^{(d)})$ &1.71&1.66&2.30&{\textbf{2.39}}&1.91&1.71\\ \hline
			$M(G_{{tv}_1}^{(a)},G_{{tv}_2}^{(a)})$ &1.78&1.70&2.36&{\textbf{2.58}}&1.92&1.83\\ \hline
			$M(G_{{tv}_1}^{(a)},G_{{tv}_3}^{(a)})$ &2.41&2.28&3.59&{\textbf{3.81}}&2.76&2.49\\ \hline
			$M(G_{{tv}_2}^{(a)},G_{{tv}_3}^{(a)})$ &1.73&1.68&2.28&{\textbf{2.47}}&1.91&1.79\\ \hline\hline
			\Xhline{1.2pt}
		\end{tabular}
		\vspace{-1em}
	\end{table}
	
	\subsection{Fast determination for the ternary tree-based partition}
	
	The TT-based partition will divide a block into three parts and calculate the R-D cost for each of them, which makes the partition more complicated than the BT-based one. Compared with the BT-based splitting, fewer TT-based partition modes will be attempted in VVC coding\cite{12}. Therefore, to reduce the complexity as much as possible, we decide how to skip the TT-based splitting modes with fewer restrictions than the BT-based modes.
	
	It is found from Table \ref{VGR} that THS has the largest difference when the gradient direction is vertical. It indicates that the vertical difference is more related to the THS compared with the other gradient-based differences. Similarly, when the gradient direction is horizontal, the TVS has the largest difference, as presented in Table \ref{135GR}. As a result, the horizontal gradient difference is more related to the TVS compared with the other gradient differences. Based on this fact, we determine the skip flags for THS and TVS as
	
	\vspace{-.5em}
	\begin{eqnarray} \label{Eq.7}
	{skip}_{th}=
	\left\{
	\begin{array}{rcl}
	\begin{aligned}
	&1, if {\left(\begin{subarray}\\
		M(G_{th_1}^{(v)},G_{th_2}^{(v)}) < T_3\\
		M(G_{th_1}^{(v)},G_{th_3}^{(v)}) < T_3\\
		M(G_{th_2}^{(v)},G_{th_3}^{(v)}) < T_3
		\end{subarray}\right)}\\
	&0, otherwise\\
	\end{aligned}
	\end{array} \right.
	\end{eqnarray}
	
	\noindent
	and
	
	\vspace{-1em}
	\begin{eqnarray} \label{Eq.8}
	{skip}_{tv}=
	\left\{
	\begin{array}{rcl}
	\begin{aligned}
	&1, if {\left(\begin{subarray}\\
		M(G_{tv_1}^{(h)},G_{tv_2}^{(h)}) < T_3\\
		M(G_{tv_1}^{(h)},G_{tv_3}^{(h)}) < T_3\\
		M(G_{tv_2}^{(h)},G_{tv_3}^{(h)}) < T_3
		\end{subarray}\right)}\\
	&0, otherwise\\
	\end{aligned}
	\end{array}, \right.
	\end{eqnarray}
	
	\noindent
	where $T_3$ is the pre-defined threshold, and ${skip}_{th}=1$ and ${skip}_{tv}=1$ indicate skipping the THS mode and the TVS mode, respectively. Based on Eqs. (\ref{Eq.7}) and (\ref{Eq.8}), if the gradients of two sub-blocks are similar along a specific direction, the block will not be split along this direction.
	
	\begin{algorithm}[t]
		\small
		{
			\caption{Proposed binary tree-based partition}
			\label{Alg.1}
			\textbf{Input:} $G_{i}^{(h)}$, $G_{i}^{(v)}$, $G_{i}^{(a)}$, $G_{i}^{(d)}$($i=$  $bh_1$,$bh_2$,$bv_1$,$bv_2$)\\
			~~~~~~~~~~~~~and $D(j,k)$($j,k=$  $bh_1$,$bh_2$,$bv_1$,$bv_2$).
			
			\textbf{Output:} ${skip}_{bh}$, ${skip}_{bv}$.\\
			\textbf{Initialization:} ${skip}_{bh}=0$, ${skip}_{bv}=0$.\\
			Calculate ${d}^{(g)}_{bh}$ according to Eq.(\ref{Eq.2});\\
			\If {${d}^{(g)}_{bh} = 0$}
			{{${skip}_{bh} = 1$.\\}
				\Else{Calculate ${d}^{(c)}_{bv}$ according to Eq.(\ref{Eq.6});\\
					\If{${d}^{(c)}_{bv} = 1$}{${skip}_{bh} = 1$.}
				}
			}
			
			Calculate ${d}^{(g)}_{bv}$ according to Eq.(\ref{Eq.3});	\\		
			\If {${d}^{(g)}_{bv} = 0$}
			{${skip}_{bv} = 1$.\\
				\Else{Calculate ${d}^{(c)}_{bh}$ according to Eq.(\ref{Eq.5});\\
					\If{${d}^{(c)}_{bh} = 1$}{${skip}_{bv} = 1$.}
				}
			}
		}
	\end{algorithm}
	
	\section{Experimental Results}
	
	We apply our proposed method to the coding of luminance (Y) component in VVC and verify its effectiveness on both VTM-5.0 and VTM-9.3. VTM-5.0 is employed in the simulations so that we can make a fair comparison with the adaptive CU split decision (ADSD) method ~\cite{6} which is implemented based on VTM-5.0. VTM-9.3 is adopted in our work because we developed our method based on it. The AI configuration is used in this work and QPs are specified as $\{22, 27, 32, 37\}$. We compress 100 frames of each test video sequence from class A1 to E. Compared with VVC, the BD-rate saving and the encoding time saving (TS) are employed to evaluate the R-D performance and the complexity, respectively. Meanwhile, the thresholds $\{T_1, T_2, T_3\}$  used in our method are empirically specified as $\{1.165, 3.500, 2.000\}$ and $\{1.180, 3.500, 2.000\}$ for VTM-5.0 and VTM-9.3, respectively.
	
	We firstly present some experimental results in Table \ref{result1} to compare our method with the ADSD method~\cite{6}. To make a fair comparison, we only offer the results for the same test video sequences used in ~\cite{6}. One can see from Table \ref{result1} that our proposed method saves more time and achieves better R-D performance than the ADSD method. We integrate our proposed method in VTM-9.3 to make a comprehensive verification of its effectiveness. The corresponding results are presented in Table \ref{result3}. Comparing the results presented in Tables \ref{result1} and \ref{result3}, one can find that integrating our proposed method in VTM-5.0 and VTM-9.3  achieves similar performances, which demonstrates the effectiveness and robustness of our proposed method.
	
	\begin{table}[t]
		\vspace{-1.5em}
		\renewcommand\arraystretch{1.2}
		\centering
		\fontsize{7}{7}\selectfont
		\caption{Comparison with ADSD~\cite{6}}\label{result1}
		\vspace{-.5em}{
			\begin{tabular} {p{5pt} p{50pt} c c c c }
				\Xhline{1.2pt}
				\hline\hline
				\multicolumn{2}{c}{\multirow{2}{*}{}}&\multicolumn{2}{c}{Ours}& \multicolumn{2}{c}{ADSD~\cite{6}}\\
				\cline{3-6}
				\multicolumn{2}{c}{}&BD-rate & TS & BD-rate & TS\\
				\hline
				A&Campfire &1.03\% &47.47\% &1.05\% &34.96\% \\
				\hline
				&Kimono &0.25\% &41.82\% &0.87\% &33.32\%  \\
				B &ParkScene &0.90\% &35.30\% &0.83\% &35.41\%  \\
				&BQTerrace &1.14\% &39.48\% &0.95\% &34.50\%   \\
				\hline
				&PartyScene &0.64\% &32.72\% &0.55\% &31.10\%  \\
				C &RaceHorses &0.88\% &39.39\% &0.37\% &23.63\%  \\
				&BasketballDrill &2.06\% &38.57\% &1.30\% &33.39\%  \\
				\hline
				&BlowingBubbles &0.81\% &32.13\% &0.95\% &33.90\% \\
				D  &RaceHorses &0.75\% &33.70\% &0.71\% &31.79\%\\
				&BQSquare &0.75\% &26.47\% &0.68\% &30.73\% \\
				\hline
				&Video1 &1.06\% &35.44\% &1.63\% &38.73\%\\
				E &FourPeople &1.03\% &33.74\% &1.38\% &38.01\%\\
				&KristenAndSara &1.12\% &37.40\% &1.61\% &34.84\%\\
				\hline
				&\textbf{Average}  &\textbf{0.96}\% &\textbf{37.13}\% &0.99\% &33.41\% \\
				\hline\hline
				\Xhline{1.2pt}
		\end{tabular}}
		\label{tab:Margin_settings}
		\vspace{-1em}
	\end{table}
	\begin{table}[t]
		\caption{Results obtained based on VTM-9.3}\label{result3}
		\vspace{-.5em}
		\renewcommand\arraystretch{1.2}
		\centering
		\fontsize{7}{7}\selectfont
		{
			\begin{tabular}{p{5pt}p{50pt}ccccc}
				\Xhline{1.2pt}
				\hline\hline
				\multicolumn{2}{c}{}&YUV &Y &U &V  & TS\\
				\hline
				\multirow{3}{*}{A1}
				&Tango2&0.82\%&0.85\%&0.24\%&0.43\%&59.54\%\\
				&FoodMarket4 &0.19\%&0.25\%&0.03\%&-0.01\%&49.60\%\\
				&Campfire&1.13\%&1.43\%&0.16\%&0.39\%&49.82\%\\
				\hline
				\multirow{3}{*}{A2}
				&CatRobot1 &1.04\%&1.10\%&0.62\%&0.69\%&43.85\%\\
				&DaylightRoad2&1.42\%&1.42\%&0.90\%&0.83\%&52.68\%\\
				&ParkRunning3 &0.35\%&0.37\%&0.32\%&0.34\%&39.66\%\\
				\hline
				\multirow{5}{*}{B}
				&MarketPlace &0.80\%&0.84\%&0.55\%&0.42\%&51.23\%\\
				&RitualDance &0.94\%&1.01\%&0.65\%&0.68\%&33.95\%\\
				&Cactus &1.04\%&1.12\%&0.55\%&0.63\%&46.03\%\\
				&BasketballDrive &1.04\%&1.11\%&0.24\%&0.47\%&49.54\%\\
				&BQTerrace &1.19\%&1.19\%&1.19\%&1.01\%&41.59\%\\
				\hline
				\multirow{4}{*}{C}
				&BasketballDrill &2.31\%&2.52\%&1.09\%&1.27\%&39.78\%\\
				&BQMall &0.91\%&0.95\%&0.46\%&0.76\%&35.37\%\\
				&PartyScene &0.61\%&0.65\%&0.34\%&0.29\%&33.12\%\\
				&RaceHorses &0.93\%&1.02\%&0.56\%&0.33\%&40.03\%\\
				\hline
				\multirow{4}{*}{D}
				&BasketballPass &0.90\%&0.97\%&0.31\%&0.66\%&36.10\%\\
				&BQSquare&0.76\%&0.79\%&0.54\%&0.28\%&29.38\%\\
				&BlowingBubbles&0.82\%&0.86\%&0.49\%&0.65\%&32.71\%\\
				&RaceHorses&0.69\%&0.76\%&0.21\%&0.45\%&34.51\%\\
				\hline
				\multirow{3}{*}{E}
				&FourPeople&1.01\%&1.08\%&0.59\%&0.48\%&34.06\%\\
				&Johnny&1.36\%&1.38\%&1.32\%&1.09\%&46.10\%\\
				&KristenAndSara &1.16\%&1.20\%&0.81\%&0.99\%&37.35\%\\
				\hline
				&\textbf{Average}  &\textbf{0.97}\% &\textbf{1.04}\% &\textbf{0.55}\% &\textbf{0.60}\% &\textbf{41.64}\% \\
				\hline\hline
				\Xhline{1.2pt}
		\end{tabular}}
		\label{tab:Margin_settings}
		\vspace{-1.5em}
	\end{table}

	\section{Concluding Remarks}
	
	In this paper, we propose a new fast CU partition algorithm for VVC intra coding based on the cross-block difference which is measured by the gradient difference and the content difference of sub-blocks belonging to a separated block. This cross-block difference is used to guide the skipping of unnecessary partition modes, which accordingly leads to a fast determination for CU partition. Compared with VVC, our proposed method save 41.64\% (on average) encoding time with only 0.97\% (on average) increase of BD-rate.

\end{document}